\documentclass[sigconf,balance=false]{acmart}
\usepackage{amsfonts}
\usepackage{amssymb}
\usepackage{multirow}
\usepackage{enumitem}
\usepackage{subcaption}
\usepackage{pifont}
\usepackage{algorithm}
\usepackage[noend]{algpseudocode}
\DeclareMathOperator*{\argmax}{arg\,max}
\newif\ifshowconcepts
\showconceptsfalse                                          
\newcommand{\concept}[1]{\ifshowconcepts{\color{orange}#1}\else{#1}\fi}
\usepackage{xspace}
\newcommand{\ie}{i.e.,\xspace}
\newcommand{\eg}{e.g.,\xspace}
\newcommand{\cf}{cf.\xspace}
\newcommand{\ok}{\ding{51}}
\newcommand{\no}{\ding{55}}
\newcommand{\na}{--}                                        

\setcopyright{none}
\acmYear{2027}
\copyrightyear{2027}
\acmDOI{}
\acmISBN{}
\title[RecRSI: Ranking-Scored Trials for Recommender RSI]{Try Before You Evolve: Ranking-Scored Trials for Recursive Self-Improvement of Recommender Systems}
\author{Xu Huang}
\email{xu\_huang@mail.ustc.edu.cn}
\affiliation{%
  \institution{University of Science and Technology of China}
  \city{Hefei}
  \country{China}
}

\author{Jiawei Feng}
\email{jwf3ng@mail.ustc.edu.cn}
\affiliation{%
  \institution{University of Science and Technology of China}
  \city{Hefei}
  \country{China}
}

\author{Jiancan Wu}
\email{wujcan@gmail.com}
\affiliation{%
  \institution{University of Science and Technology of China}
  \city{Hefei}
  \country{China}
}
\renewcommand{\shortauthors}{Xu Huang, Jiawei Feng, and Jiancan Wu}
\begin{document}
\begin{abstract}
Recommenders on Web platforms such as e-commerce and music streaming need their training recipes re-tuned for each dataset.
Large language model (LLM) agents can now rewrite these recipes, enabling recursive self-improvement (RSI) of recommender systems.
Existing agents propose one edit per round, fully retrain the recommender, and keep the edit only if its ranking quality improves.
Unlike programs or patches that can be scored by execution, a recommender edit changes how the model is trained, which raises two challenges.
First, ranking feedback is delayed: an edit's ranking quality is known only after full retraining, so much of the retraining budget goes to edits that are later discarded.
Second, the signals available before retraining are unreliable: sampled training losses can look solved while full-ranking performance remains poor, and they are not comparable across edits that change the loss or sampler.
We propose RecRSI, which generates diverse candidate edits from a diagnosis of the current model and selects one for full retraining using early ranking feedback: each candidate receives a short trial, the first steps of its full training run, scored by full-ranking NDCG@10 on held-out probe users.
With two sequential recommenders on three datasets and the same number of full retrainings, RecRSI significantly outperforms existing self-evolving agents in every setting, and on both Amazon datasets reaches in its first round a ranking quality that none of them reaches in five.
\end{abstract}

\maketitle
\pagestyle{plain}
\section{Introduction}
\label{sec:intro}
\begin{figure*}[t]
  \centering
  \includegraphics[width=\textwidth]{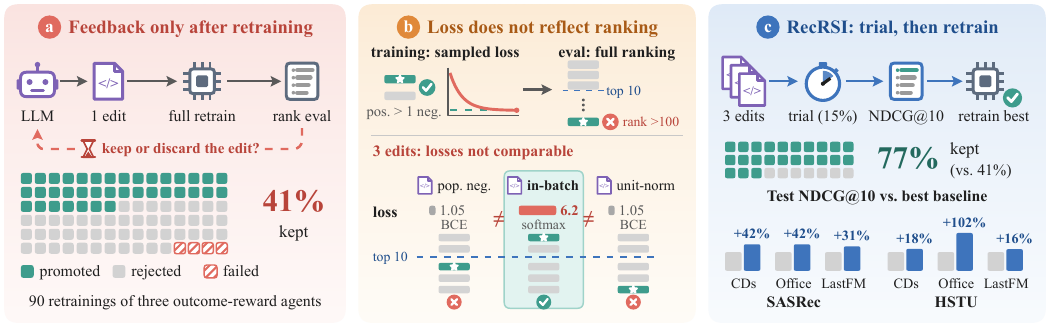}
  \vspace{-14pt}
  \caption{\textbf{Why self-improving recommenders need a signal before the retraining.} (a) Ranking feedback arrives only after a full retraining, and across three agents (AlphaEvolve, DeepEvolve and Self-EvolveRec), two backbones (SASRec and HSTU) and three datasets (CDs, Office and LastFM), 37 of 90 retrainings (41\%) yield a kept edit. (b) The sampled training loss settles at its floor while the positive ranks far below the top 10, and edits that change the loss or sampler yield incomparable losses: after a short trial, the edit with the highest loss ranks best. (c) RecRSI trains three edits briefly and retrains the one with the highest trial score, keeping 23 of 30 retrainings and outperforming the strongest existing agent (gray) in all six settings.}
  \label{fig:teaser-figure}
  \Description{Three panels. Panel a: icons of an LLM, one code edit, a GPU labelled full retrain and a ranked list labelled rank eval, connected by arrows; a red dashed arrow with an hourglass returns from the ranked list to the LLM, labelled keep? only after retraining; a unit chart of 90 squares with 37 promoted, 49 rejected and 4 failed retrainings, labelled 41 percent kept. Panel b, top: training with a sampled loss, a starred positive above one sampled negative with a check mark and a loss curve settling onto a dashed floor, followed by an arrow to full-ranking evaluation, where the same positive lies below a dashed top-10 cut at rank beyond 100 with a cross. Panel b, bottom, headed 3 edits: losses not comparable: three edits (popularity negatives, in-batch softmax, unit-norm embeddings) with training losses 1.05 (BCE), 6.2 (softmax) and 1.05 (BCE) separated by not-equal signs, and ranked lists in which the positive falls below the top-10 cut for the two BCE edits (crosses) and sits at the top for the highlighted softmax edit (check). Panel c: icons of three edits, a stopwatch, a ranked list and a GPU; a unit chart of 30 squares with 23 promoted, labelled 77 percent kept versus 41 percent; six bar pairs of Test NDCG@10 relative to the strongest baseline: plus 42, 42 and 31 percent for SASRec and plus 18, 102 and 16 percent for HSTU on CDs, Office and LastFM.}
\end{figure*}

The training recipe of a recommender, including but not limited to its architecture \cite{he2020lightgcn,kang2018sasrec,sun2019bert4rec,yang2023dreamrec,zhai2024hstu}, training loss \cite{rendle2009bpr,wu2024ssm,klenitskiy2023dross,petrov2023gsasrec}, negative sampler \cite{rendle2014improving,yi2019sampling} and optimization schedule \cite{kingma2015adam,rendle2019difficulty}, largely determines how well it performs, and the best recipe varies across domains and datasets \cite{ferrari2019troubling,zhang2020retrain}.
Consequently, identifying an effective recipe still relies on manual and iterative tuning, in which practitioners repeatedly modify the training code and retrain the model.
Large language model (LLM) agents have begun to take over part of this work.
Building on coding agents that evolve programs from execution feedback~\cite{romera2024funsearch,novikov2025alphaevolve,liu2025deepevolve}, recommender agents rewrite the training code of a recommender, retrain the edited model, and keep the edit when the retrained model ranks better on validation users~\cite{kim2026selfevolverec,wang2026selfevolving,mu2026evorec,ling2026recharness,yang2026agentx}, bringing recursive self-improvement (RSI) from general agents~\cite{zhang2025dgm,wang2025hgm} into recommender development.

This loop is inherited from domains in which candidates can be evaluated directly by execution, yet recommender edits depart from this setting in an essential respect.
Programs for algorithmic problems~\cite{romera2024funsearch,novikov2025alphaevolve} and patches for coding benchmarks~\cite{zhang2025dgm,wang2025hgm} can be assessed by executing them on test instances, whereas a recommender edit modifies the procedure through which the model is learned, so that its \concept{outcome reward}~\cite{uesato2022solving,lightman2023verify}, the validation gain of the edited recommender, can be observed only after the model has been retrained on the interaction data.
Consequently, proposing an edit requires a single LLM call, whereas evaluating it requires a complete retraining.
Under this asymmetry, an agent is able to propose many candidate edits but to retrain few of them, and the improvement it attains depends largely on which candidates are selected for retraining.
Existing agents propose one edit per round and retrain it without prior assessment, so that each retraining is spent on the first proposal of the LLM, and fewer than half of these retrainings yield a kept edit (Figure~\ref{fig:teaser-figure}(a)).
A natural alternative is to propose several edits per round and to retrain the most promising among them, provided that their relative promise can be estimated at a small fraction of the cost of a full retraining.

Such an estimate is, however, non-trivial to obtain for recommender edits, because they undermine the signals on which early evaluation conventionally relies.
In hyperparameter search, candidate configurations share a training objective and an evaluation protocol, so that their early losses or learning curves can be compared and extrapolated~\cite{li2017hyperband,domhan2015speeding}.
Recommender code edits violate this premise.
Recommenders are commonly trained with sampled objectives, and studies of sampled evaluation have shown that outscoring a small set of sampled items reveals little about the rank of an item under the full-ranking protocol~\cite{krichene2020sampled,petrov2023gsasrec}, so the training loss may approach its floor while full-ranking performance remains poor.
The difficulty is compounded when an edit targets the objective itself, as many effective edits do, since modifying the loss or the negative sampler also alters the scale on which training progress is measured and thereby deprives competing edits of a common basis for comparison (Figure~\ref{fig:teaser-figure}(b)).
Signals that assess an edit without training it, such as parent diagnosis or LLM patch judgments, avoid this difficulty but ignore how the edited model learns.

These observations suggest that an early signal should be obtained through the same procedure as the outcome reward, with a shorter training run, rather than through a different quantity.
The outcome reward is informative because it runs the edited training procedure and judges the resulting model by full-ranking performance, and its shortcoming lies in when it arrives rather than in what it measures.
An early signal that runs the same training procedure for fewer steps and applies the same full-ranking measurement, two properties we call \concept{training consistency} and \concept{ranking consistency}, preserves what makes the outcome reward reliable at a fraction of its cost.
Such brief runs resemble the low-fidelity evaluations of hyperparameter search~\cite{jamieson2016successive,li2017hyperband}, but for recommender code edits their usefulness depends on keeping both properties, which signals read from the training loss give up.

Building on this principle, we propose \textbf{RecRSI}, a framework that compares candidate edits through short trials and spends the single full retraining of each round on the most promising one.
In each round, the LLM proposes several diverse edits grounded in a diagnosis of the current model and a record of earlier rounds, and each executable edit receives a \concept{short trial}, which trains the edited recommender from the same initialization as the full retraining for a small fraction of its training steps and scores it by full-ranking NDCG@10 on held-out validation users.
The edit with the highest trial score then receives the full retraining and is kept if it improves validation performance under the same rule as in existing agents.
Because the short trial reuses the training and evaluation code of the full retraining and requires no learned reward model, this selection step can be incorporated into existing recommender agents.

We evaluate RecRSI with SASRec and HSTU on two Amazon datasets and a music listening log, granting every method the same number of full retrainings and reporting the additional cost of the trials separately.
First, RecRSI outperforms the strongest existing agent in every setting, by 16.1\% to 102.0\% in Test NDCG@10, and 23 of its 30 retrainings yield a kept edit, compared with 37 of 90 for existing agents. Under an equal one-GPU-hour budget on SASRec and CDs, it also outperforms all three agents and a batch proposer.
Second, the ablations attribute this gain to the short trial, as replacing it with a random choice among the same edits, an LLM judgment or a continuation from the current model lowers the final quality on average, as does retraining untried edits for more rounds, and executed signals track the outcome rewards of the edits more closely than non-executed ones.
Third, RecRSI keeps an edit in the first round of all six runs, and on both Amazon datasets it reaches within one round a validation NDCG@10 that no existing agent reaches within five.

\section{Preliminaries and Motivation}
\label{sec:prelim}
Recent agents~\cite{kim2026selfevolverec,wang2026selfevolving,mu2026evorec,ling2026recharness} improve a recommender by letting an LLM rewrite its training code and keeping a rewrite only if the retrained model ranks better for held-out users.
We formalize this loop, explain why its feedback is late and hard to replace, and derive two requirements for early feedback.

\subsection{Recommender Evolution Loop}
\label{sec:loop}
A recommender with parameters $\theta$ assigns a score $s_\theta(u,i)$ to every item $i$ in the item set $\mathcal{I}$ for user $u$, and its quality is judged by how highly it ranks each user's held-out item among all items of $\mathcal{I}$, which we refer to as full-ranking evaluation.
We refer to the source code that produces such a model as its \textit{recipe} $T$, which contains everything a practitioner would tune when building recommendation models like SASRec~\cite{kang2018sasrec}, including the architecture, the training loss, the negative sampler, the optimizer and its schedule, and how the model produces predictions (\eg weight averaging).
Each user's interactions are split in time into training interactions, a validation interaction and a later Test interaction, and models are trained on training interactions only.
A fixed hash of the user identifier divides the validation interactions into two disjoint sets of users:
\begin{itemize}[leftmargin=*]
    \item \textit{Probe users} $\mathcal{U}^{\text{pr}}$ serve early stopping, and our method also uses them to compare candidate edits.
    \item \textit{Select users} $\mathcal{U}^{\text{sel}}$ decide whether to promote a retrained recipe.
\end{itemize}
The Test interactions of all users are used only for the final report and never enter the search (Appendix~\ref{apdx:impl-details}).
Keeping the two validation sets apart prevents the repeated comparisons on probe users from overfitting the decision of which recipe to keep.
To compare recipes on equal terms, all recipes are trained under one fixed training setup, in which the initial parameters $\theta^{(0)}$ are drawn with a shared seed $\xi$ and training runs under a predefined step budget with early stopping on probe users.
Let $\theta_T$ denote the parameters learned by fully training recipe $T$, and let $Y(T)$ be its full-ranking NDCG@10 on select users.

An agent improves the recipe over rounds.
At round $t$, the current recipe $T_p$, which we call the \textit{parent}, has the trained checkpoint $\theta_p=\theta_{T_p}$.
An LLM reads $T_p$ and proposes a set of edits $\mathcal{A}_t$, where each edit $a$ is a code change, for instance replacing the binary cross-entropy loss of SASRec with an in-batch softmax, that turns $T_p$ into a new recipe $T_a$.
The agent learns whether an edit helps only through its \concept{outcome reward}
\begin{equation}
  \Delta Y(a)=Y(T_a)-Y(T_p),
  \label{eq:outcome}
\end{equation}
which is the validation gain that a full retraining of $T_a$ reveals.
Since retraining is expensive, every agent receives the same budget of one full retraining per round, and the retrained recipe replaces the parent, which we call \textit{promotion}, only if $\Delta Y(a)>\delta$ for a fixed margin $\delta$.
Existing agents propose a single edit per round ($|\mathcal{A}_t|=1$), so an edit is retrained as soon as it is proposed~\cite{novikov2025alphaevolve,liu2025deepevolve,kim2026selfevolverec}.
We study the general case $|\mathcal{A}_t|=K$, in which the agent must first decide which of the $K$ edits deserves the retraining by means of a signal $S(a)$ that is much cheaper than a retraining.

\subsection{Why Choosing What to Retrain Is Hard}
\label{sec:motivation}
We examine this difficulty with AlphaEvolve, DeepEvolve and Self-EvolveRec, adapted to a common recommender harness (\cf Section~\ref{sec:setup}) and run for five rounds from the same initial models.

\textbf{Feedback arrives only after retraining.}
An edit changes how the model is trained, so the outcome reward that decides whether it is kept is measured after a full training run has finished.
Each retraining trains the edited recipe on the whole interaction log until early stopping, and this cost grows with the item catalog and the interaction log (Appendix~\ref{apdx:cost} reports the measured times).
Since these agents propose one edit per round, each retraining is spent on an edit that has never been run, and most of these retrainings do not lead to a kept edit.
Across the three frameworks, only 37 of the 90 retrainings yield an edit that is kept (11, 11 and 15 of 30 for AlphaEvolve, DeepEvolve and Self-EvolveRec, Figure~\ref{fig:teaser-figure}(a)), and four retrainings of Self-EvolveRec fail before training starts.
Progress is also slow, as even Self-EvolveRec, which diagnoses the parent before every proposal, keeps its first SASRec edit only in round 3 on CDs and in round 4 on Office.

\textbf{Training loss does not reflect ranking.}
A natural way to avoid this cost is to judge an edit by a signal available early in training, such as its training loss, as multi-fidelity search does~\cite{li2017hyperband,domhan2015speeding}.
SASRec is trained with binary cross-entropy, which only requires the positive item to score above one sampled negative, whereas NDCG@10 requires it to score above almost every item in the catalog.
This is the same gap that makes sampled evaluation unreliable~\cite{krichene2020sampled,petrov2023gsasrec}.
On Office, only 6.4\% of positives score below their sampled negative under the initial SASRec model, so its training loss is close to its floor, yet 55\% of positives score below the strongest of just 32 random items (Figure~\ref{fig:teaser-figure}(b)), indicating that more than half of the positives rank below position 100.\footnote{A positive at rank $r$ in a catalog of $N$ items scores below the strongest of 32 random items with probability about $1-(1-r/N)^{32}$, which is about 4\% for $r=100$ and $N=77{,}551$, so a 55\% rate implies that at least 53\% of positives rank below 100.}

\begin{figure*}[t]
  \centering
  \includegraphics[width=\textwidth]{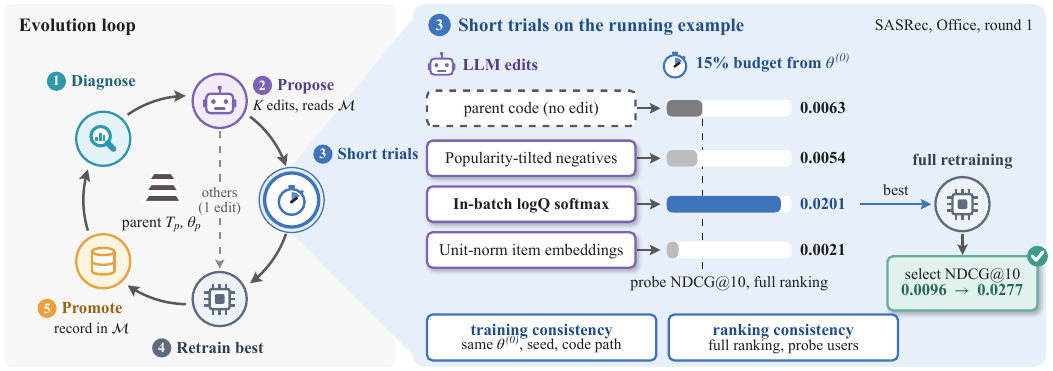}
  \vspace{-16pt}
  \caption{\textbf{The evolution loop of RecRSI (left) and its short-trial step on the running example (right).} In each round, RecRSI (1) diagnoses the parent on probe users, (2) asks the LLM for $K$ edits given the diagnosis and the memory $\mathcal{M}$, (3) trains each valid edit and the parent recipe from $\theta^{(0)}$ for the first 15\% of the step budget of its full training run and scores them by full-ranking NDCG@10 on probe users, (4) spends the full retraining on the edit with the highest trial score, and (5) promotes it if select NDCG@10 improves by more than $\delta$ and records the round in $\mathcal{M}$. Existing agents propose a single edit and skip step (3). Right: measured trial scores for SASRec on Office in round 1, where the dashed line marks the parent recipe and two of the three edits fall below it.}
  \label{fig:main-figure}
  \Description{Left: a circular loop of five icons around a stacked-layer parent model: (1) diagnose with a magnifier, (2) propose with an LLM robot, (3) short trials with a stopwatch, highlighted and magnified to the right, (4) retrain best with a GPU chip, and (5) promote with a memory cylinder. Right: the parent recipe and three edits named popularity-tilted negatives, in-batch logQ softmax and unit-norm item embeddings, each with a bar of probe NDCG@10 after training for 15 percent of the step budget from theta zero: 0.0063, 0.0054, 0.0201 and 0.0021; the best edit goes to a full retraining that raises select NDCG@10 from 0.0096 to 0.0277; two boxes name training consistency and ranking consistency.}
  \vspace{-4pt}
\end{figure*}

The gap is especially relevant to the edits that help.
On Office, five of the seven SASRec edits kept by the three frameworks change the loss or the negative sampler.
Once an edit changes the loss, its loss values are on a different scale from those of other edits and are not directly comparable.
In the first round on Office (Figure~\ref{fig:main-figure}, right), after training for 15\% of the step budget, the edit that switches to an in-batch softmax reaches a loss of 6.2, against about 1.05 for the two edits that keep binary cross-entropy, so choosing the edit with the lowest training loss would discard it, although it is the one of the three that ranks above the parent.

\begin{table}[t]
  \centering
  \caption{Signals for choosing which recommender edit to retrain. Run: executes recommender code on data; Candidate: assesses the proposed edit itself; Train.: follows the edit\textquotesingle s full-training procedure from the shared initialization; Rank: measures full-ranking NDCG@10 on held-out users; Early: available before full retraining. A checkmark indicates that the signal has the property.}
  \label{tab:signals}
  \footnotesize
  \setlength{\tabcolsep}{1.5pt}
  \begin{tabular}{l|ccccc|l}
    \toprule
    \textbf{Signal} & \textbf{Run} & \textbf{Candidate} & \textbf{Train.} & \textbf{Rank} & \textbf{Early} & \textbf{Examples}\\
    \midrule
    Outcome reward & \ok & \ok & \ok & \ok & \no & AlphaEvolve~\cite{novikov2025alphaevolve}\\
    Subset evaluation & \ok & \ok & \ok & \ok & \no & DGM~\cite{zhang2025dgm}, HGM~\cite{wang2025hgm}\\
    Parent diagnosis & \ok & \no & \na & \ok & \ok & Self-EvolveRec~\cite{kim2026selfevolverec}\\
    LLM patch judge & \no & \ok & \no & \no & \ok & SIFT~\cite{fu2026sift}\\
    Zero-cost proxy & \ok & \ok & \no & \no & \ok & NAS~\cite{abdelfattah2021zerocost,mellor2021naswot}\\
    Curve extrapolation & \ok & \ok & \ok & \no & \ok & \citet{domhan2015speeding}\\
    Probe at parent & \ok & \ok & \no & \ok & \ok & warm start (\S\ref{sec:rq2})\\
    \midrule
    Short trial & \ok & \ok & \ok & \ok & \ok & RecRSI (ours)\\
    \bottomrule
  \end{tabular}
  \vspace{-8pt}
\end{table}

\subsection{Requirements for a Pre-Retraining Signal}
\label{sec:requirements}
The outcome reward falls short in its timing rather than in what it measures, since it retrains the edited recipe from $\theta^{(0)}$ and scores it by full-ranking evaluation, but does so after the retraining has been spent.
A cheaper signal should therefore keep both of these parts, which we state as two properties:
\begin{itemize}[leftmargin=*]
  \item \textbf{Training consistency.}
    The signal is computed from a model obtained by training $T_a$ as the full retraining does, from $\theta^{(0)}$ with seed $\xi$ and with the edit's own architecture, loss, sampler, optimizer and schedule, possibly for fewer steps.
  A signal that does not run this training cannot see edits that act through the loss or the sampler, which make up most kept edits, and continuing from $\theta_p$ instead changes the training path being evaluated.
  \item \textbf{Ranking consistency.}
  The signal evaluates the trained model using full-ranking NDCG@10 on held-out users, as does the outcome reward. Training loss and sampled metrics may look solved despite poor ranking and are not comparable across edits.
\end{itemize}


In Table~\ref{tab:signals}, parent diagnosis executes the current model but does not assess a candidate edit, whereas an LLM patch judge assesses the candidate without executing it. Every existing early signal gives up training or ranking consistency, while signals that keep both wait for full training.
A short training run of the edit from $\theta^{(0)}$, scored by full-ranking NDCG@10, keeps both properties at a fraction of the cost, in the same way as the partial training runs of multi-fidelity hyperparameter search~\cite{jamieson2016successive,li2017hyperband}, and its remaining error comes from stopping early, which Section~\ref{sec:rq2} measures.
Together, these properties connect candidate selection to the ranking metric used for promotion. Each score reflects the learning behavior of an edited recipe, providing a common basis for comparing changes to different training components.

\section{RecRSI}
\label{sec:method}
In this section, we elaborate on the proposed RecRSI, as illustrated in Figure~\ref{fig:main-figure}, which builds each round of the evolution loop around the short training run identified in Section~\ref{sec:requirements}, referred to as a \concept{short trial}.
A short trial trains an edited recipe from $\theta^{(0)}$ for the first $K_s$ steps of its full training run and scores it by full-ranking NDCG@10 on probe users, and this score serves as the signal $S(a)$ that decides which of the proposed edits receives the round's full retraining.
Each round still spends one full retraining, which now goes to the edit with the highest trial score among several tried edits rather than to a single untried one, at the additional cost of the trials (Figure~\ref{fig:main-figure}).
We first describe how RecRSI proposes a batch of edits (Section~\ref{sec:proposal}), then specify the short trial (Section~\ref{sec:trial}), and finally describe how the round spends its retraining and records what it has learned (Section~\ref{sec:commit}).

\subsection{Diagnosis-Conditioned Batch Proposal}
\label{sec:proposal}
The benefit of selecting among edits depends on the batch from which the selection is made, which should consist of edits that address the failures of the parent and that differ from one another, since near-identical edits are likely to yield similar outcomes.
To this end, RecRSI grounds the proposal in a diagnosis of the parent and asks the LLM to distribute its edits over different components of the recipe.
 
The diagnosis is computed by a forward pass of $\theta_p$ over probe users and summarizes where the ranking of the parent fails.
It includes the rate from Section~\ref{sec:motivation} at which a positive scores below the strongest of 32 random items, together with the average score margin to that item, and the share of score variance explained by item identity alone, which indicates how close the parent is to ranking items in the same order for every user.
It also reports how the positives of probe users are distributed over ranks 1--10, 11--100 and beyond 100 under the parent, as well as NDCG@10 on head and tail items, so that each hypothesis can name the group it aims to improve.
The diagnosis describes the parent rather than any edit, and RecRSI therefore uses it to inform which edits are proposed while leaving the choice among them to the short trial.
 
The prompt combines the diagnosis with the source code of the parent, the interface between a recipe and the training harness, and the memory $\mathcal{M}$ of earlier rounds (Section~\ref{sec:commit}).
To keep the edits distinct, the LLM is asked to place them on different components of the recipe, such as the loss, the negative sampler, the optimizer, the prediction and the architecture, so that the batch is more likely to cover several hypotheses than variants of one.
Each edit is returned as search-and-replace patches together with the component it changes, a hypothesis that cites the diagnosis, and the rank group it expects to improve first.
 
Since edits are free-form code, RecRSI checks each of them before it touches real data, both to avoid spending a trial on code that cannot run and to reduce the risk that an edit alters how models are evaluated.
The evaluation code is kept fixed and outside the editable recipe, and an edit is accepted when its patches apply cleanly, a static check finds no disallowed imports or attribute access, and a smoke test on a synthetic batch shows that the edited recipe completes a training step and still exposes the scoring interface that the evaluator calls.
An edit that fails these checks receives one repair request containing the error message and is dropped if it fails again, and an edit whose code is identical to a recipe trained earlier in the run is removed as a duplicate.
The remaining edits form the candidate set $\mathcal{A}_t$, with $|\mathcal{A}_t|\le K$.
 
\subsection{Training and Ranking Consistency}
\label{sec:trial}
For each candidate $a\in\mathcal{A}_t$, the short trial yields the score
\begin{equation}
  m_a=\mathrm{NDCG@10}\bigl(\theta^{(K_s)}_{a};\,\mathcal{U}^{\text{pr}}\bigr),\qquad \theta^{(K_s)}_{a}=\mathrm{Train}_{K_s}\bigl(T_a;\theta^{(0)},\xi\bigr),
  \label{eq:trial}
\end{equation}
where $\mathrm{Train}_{K_s}$ denotes the first $K_s$ steps of the full training run without early stopping.
The parent recipe receives the same trial, which gives the reference score $m_p$ at the same training budget.
 
\textbf{Training consistency.}
The trial runs the edited recipe itself, with its own architecture, loss, sampler, optimizer and schedule, from the initialization $\theta^{(0)}$ and seed $\xi$ of the full retraining, and it follows that schedule as in the full run rather than compressing it to $K_s$ steps.
Under deterministic execution, the trial model therefore coincides with the state that the full retraining of $T_a$ reaches after $K_s$ steps, so the trial and the outcome reward observe the same training process at two points in time.
Starting from $\theta^{(0)}$ also removes the need to transfer weights from the parent, so edits that add parameters, such as a new architectural module or a learned prediction head, are evaluated alongside loss and sampler edits.
 
\textbf{Ranking consistency.}
The trial score is computed by the evaluator of the outcome reward, with the same item set, masking of each user's history and tie handling, so $m_a$ is the full-ranking NDCG@10 that defines $Y$, computed on probe users instead of select users, and it is comparable across edits regardless of the loss they train with.
Computing it on probe users keeps the select users that decide promotion out of the choice among edits. Thus, binary cross-entropy, in-batch softmax and sampler edits share a ranking metric despite different loss scales.
 
\textbf{Cost.}
A trial costs $K_s$ training steps and one evaluation on probe users, and we fix $K_s=4{,}000$, about 15\% of the step budget, in all settings.
In wall-clock time, a trial takes a median of 81--103\,s and a full retraining a median of 236--788\,s, depending on the dataset and backbone, so a round with $K$ proposed edits adds up to $K$ trials to its retraining, an overhead we report with the results (Appendix~\ref{apdx:cost}).
 
\subsection{Retraining, Promotion, and Memory}
\label{sec:commit}
RecRSI spends the round's full retraining on the candidate with the highest trial score,
\begin{equation}
  a^{\star}=\argmax_{a\in\mathcal{A}_t} m_a,
  \label{eq:select}
\end{equation}
and promotes the retrained recipe $T_{a^{\star}}$ under the rule of Section~\ref{sec:loop}, \ie if $\Delta Y(a^{\star})>\delta$.
The retraining takes place even when the highest trial score does not exceed that of the parent ($m_{a^{\star}}\le m_p$), so that every method spends one full retraining per round, and Section~\ref{sec:rq3} examines how this comparison could be used to skip retrainings.
 
After the retraining, RecRSI appends the round to the memory $\mathcal{M}$, which records each proposed edit with the component it changes and its hypothesis, the trial scores of the candidates, the promotion decision, and the change in NDCG@10 of the retrained model on each rank and popularity group of probe users.
Apart from the binary promotion decision, no select-user metric enters $\mathcal{M}$, and Test interactions stay outside the search as described in Section~\ref{sec:loop}.
The memory thus allows later proposals to avoid mechanisms whose trials fell behind and to build on those that improved the groups they targeted, without a learned model of edit quality. Even candidates that are not fully retrained contribute their trial scores to subsequent proposals, so each round gathers evidence about several mechanisms while spending one full retraining.
Algorithm~\ref{alg:recrsi} summarizes one round.
 
\begin{algorithm}[t]
  \caption{One round of RecRSI}
  \label{alg:recrsi}
  \small
  \begin{algorithmic}[1]
    \Require parent $(T_p,\theta_p)$, memory $\mathcal{M}$, batch size $K$, trial length $K_s$, margin $\delta$
    \State diagnose $\theta_p$ on probe users \Comment{forward pass only}
    \State LLM proposes $K$ edits from $(T_p$, diagnosis, $\mathcal{M})$
    \State check each edit; repair once; keep valid edits as $\mathcal{A}_t$
    \State $m_a\gets$ short trial of $T_a$ for $a\in\mathcal{A}_t$, and $m_p$ for $T_p$ \Comment{Eq.~\ref{eq:trial}}
    \State $a^{\star}\gets\argmax_{a}m_a$ \Comment{Eq.~\ref{eq:select}}
    \State retrain $T_{a^{\star}}$ from $\theta^{(0)}$; promote if $\Delta Y(a^{\star})>\delta$
    \State append edits, trial scores, decision and group changes to $\mathcal{M}$
  \end{algorithmic}
\end{algorithm}

\section{Experiments}
\label{sec:experiments}
We evaluate RecRSI through the following research questions:
\begin{itemize}[leftmargin=*,topsep=2pt,itemsep=1pt]
  \item \textbf{RQ1:} Does RecRSI improve final recommendation quality with the same number of full retrainings?
  \item \textbf{RQ2:} Does the gain come from the short-trial signal, and how do other ways of choosing among the same edits compare?
  \item \textbf{RQ3:} How do short trials affect evolution dynamics and cost?
\end{itemize}

\begin{table*}[t]
  \centering
  \caption{Final Test results after five rounds with one full retraining per round. Acc.: rounds whose retrained edit was promoted. Best results per backbone are in bold, second best are underlined. Improv.: relative gain of RecRSI over the strongest baseline in each column. }
  \label{tab:main}
  \small
  \setlength{\tabcolsep}{3.6pt}
  \begin{tabular}{ll|ccc|ccc|ccc}
    \toprule
    & & \multicolumn{3}{c|}{\textbf{CDs}} & \multicolumn{3}{c|}{\textbf{Office}} & \multicolumn{3}{c}{\textbf{LastFM}}\\
    \textbf{Backbone} & \textbf{Method} & \textbf{NDCG@10} & \textbf{HR@10} & \textbf{Acc.} & \textbf{NDCG@10} & \textbf{HR@10} & \textbf{Acc.} & \textbf{NDCG@10} & \textbf{HR@10} & \textbf{Acc.}\\
    \midrule
    \multirow{6}{*}{SASRec} & Initial & 0.0349 & 0.0667 & \na & 0.0079 & 0.0153 & \na & 0.0653 & 0.0721 & \na\\
    & AlphaEvolve & 0.0387 & 0.0701 & 2/5 & 0.0129 & 0.0211 & 3/5 & 0.0777 & 0.0873 & 3/5\\
    & DeepEvolve & \underline{0.0419} & 0.0733 & 3/5 & 0.0090 & 0.0173 & 2/5 & 0.0654 & 0.0843 & 1/5\\
    & Self-EvolveRec & 0.0418 & \underline{0.0751} & 2/5 & \underline{0.0171} & \underline{0.0310} & 2/5 & \underline{0.0915} & \underline{0.0934} & 2/5\\
    & \textbf{RecRSI (ours)} & \textbf{0.0594} & \textbf{0.0973} & 4/5 & \textbf{0.0242} & \textbf{0.0404} & 3/5 & \textbf{0.1196} & \textbf{0.1310} & 5/5\\
    & Improv. & +41.7\% & +29.5\% & & +41.5\% & +30.4\% & & +30.7\% & +40.2\% & \\
    \midrule\midrule
    \multirow{6}{*}{HSTU} & Initial & 0.0375 & 0.0697 & \na & 0.0070 & 0.0135 & \na & 0.0587 & 0.0640 & \na\\
    & AlphaEvolve & \underline{0.0476} & 0.0819 & 1/5 & \underline{0.0099} & \underline{0.0176} & 2/5 & 0.0587 & 0.0640 & 0/5\\
    & DeepEvolve & 0.0467 & \underline{0.0842} & 3/5 & 0.0074 & 0.0142 & 1/5 & 0.0586 & 0.0640 & 1/5\\
    & Self-EvolveRec & 0.0471 & 0.0835 & 2/5 & 0.0082 & 0.0156 & 3/5 & \underline{0.0742} & \underline{0.0843} & 4/5\\
    & \textbf{RecRSI (ours)} & \textbf{0.0559} & \textbf{0.0903} & 4/5 & \textbf{0.0199} & \textbf{0.0342} & 2/5 & \textbf{0.0861} & \textbf{0.0934} & 5/5\\
    & Improv. & +17.6\% & +7.2\% & & +102.0\% & +94.7\% & & +16.1\% & +10.8\% & \\
    \bottomrule
  \end{tabular}
  \vspace{-6pt}
\end{table*}

\subsection{Experimental Setup}
\label{sec:setup}

\subsubsection{Datasets.}
We use three datasets that differ in domain, scale and sequence length: CDs in the version and split released with Self-EvolveRec~\cite{kim2026selfevolverec} (14,335 users, 11,436 items), Office from the official Amazon Reviews 2023 rating-only 5-core release~\cite{hou2024blair} (223,308 users, 77,551 items), and LastFM, built from the Last.fm-1K listening log~\cite{celma2010music} (985 users, 175,635 items, 12.8M listening events).
CDs keeps its released temporal split; on Office and LastFM, each user's last interaction is held out for Test and the second-to-last for validation.
Validation interactions are divided by a fixed user hash into probe and select users, while Test covers all users on a later interaction (Section~\ref{sec:loop}).
The item catalog is built from training interactions only, and a held-out item outside it counts as a miss.
Appendix~\ref{apdx:impl-details} gives the full statistics.

\subsubsection{Backbones.}
We evolve the base models of the SASRec~\cite{kang2018sasrec} and HSTU~\cite{zhai2024hstu} architectures.
Both start from the same recipe (hidden size 50, two layers, two heads, history length 50, binary cross-entropy with one uniform negative), which the agents may rewrite.

\subsubsection{Training and evaluation.}
Every trajectory runs five rounds with exactly one full retraining per round.
A full retraining runs at most 26,000 steps from the shared initialization, evaluates probe users every 2,000 steps, and stops early after four evaluations without improvement (at least 8,000 steps).
An edit is promoted if its select NDCG@10 exceeds that of the parent by $\delta=10^{-4}$.
Within a backbone, all methods share the data split, the initial checkpoint, the seed, and the training setup.
After five rounds, the retained model is evaluated once on Test users; Test values never enter the search.
We report full-ranking NDCG@10 and HR@10, with seen items masked and ties broken by item index.
We compare methods by paired bootstrap (5,000 Test-user resamples, 95\% intervals).

\subsubsection{Baselines.}
We compare with three LLM-driven code-evolution frameworks adapted to the same harness.
AlphaEvolve~\cite{novikov2025alphaevolve} keeps an archive of scored programs and samples parents and inspirations from it; DeepEvolve~\cite{liu2025deepevolve} adds literature retrieval and reflection; and Self-EvolveRec~\cite{kim2026selfevolverec} adds numerical diagnosis of the parent, critiques from simulated users, and co-evolution of its diagnostic code.
All methods use the same open-weight LLM, GLM-5.3, the same editable files and interface checks, and the same promotion rule (Appendix~\ref{apdx:adaptation}).

\subsubsection{Implementation.}
RecRSI requests $K=3$ edits per round, allows one repair request, and runs short trials of 15\% of the full-training step budget (Appendix~\ref{apdx:impl-details}) on all 4,043 probe users of CDs, on a fixed subset of 12,000 probe users of Office, and on the 244 probe users of LastFM whose held-out item lies in the training catalog.
Appendix~\ref{apdx:cost} reports hardware and measured runtime.

\subsection{(RQ1) Final Recommendation Quality}
\label{sec:rq1}
Table~\ref{tab:main} reports the final Test quality of RecRSI and of the outcome-reward frameworks, from which we observe:

\textbf{RecRSI outperforms every baseline on both backbones and all three datasets.}
Specifically, with SASRec, RecRSI improves Test NDCG@10 over the strongest baseline by 41.7\% on CDs, 41.5\% on Office and 30.7\% on LastFM, and HR@10 by 29.5\%, 30.4\% and 40.2\%.
Moreover, with HSTU, the gains are 17.6\%, 102.0\% and 16.1\% in NDCG@10, and 7.2\%, 94.7\% and 10.8\% in HR@10, and all 36 paired comparisons with a baseline are significant.
In contrast, no baseline is consistently second: DeepEvolve is strong on CDs but weak on Office and LastFM, while Self-EvolveRec leads the baselines on SASRec Office and on LastFM but not on HSTU Office.
We attribute this robustness to the fact that RecRSI does not depend on the proposer being right in a given round, since the trials filter out the proposals that are wrong.

\textbf{The advantage is largest where the training loss hides the most ranking failures.}
Specifically, with HSTU, the margin of RecRSI over the strongest baseline grows from 17.6\% on CDs to 102.0\% on Office, where the positives of the initial HSTU lose to the strongest of 32 random items far more often (48.1\% vs.\ 7.5\% of training positions).
Such results suggest that the trials are most valuable when the training objective hides most of the ranking failure, since the training signals then tell the baselines least about which edit to keep.

\textbf{RecRSI promotes more of its retrainings.}
Across the six settings, RecRSI promotes 23 of its 30 retrained edits (77\%), including all ten on LastFM, whereas the baselines in Table~\ref{tab:main} promote 37 of 90 (41\%).
The trials thus raise the share of retrainings spent on kept edits, rather than finding a single lucky edit.
The gap is largest on LastFM, where the baselines promote only 11 of their 30 retrainings.

\begin{table}[b]
  \centering
  \caption{Equal budget of one GPU hour, trials included (SASRec, CDs, Test).}
  \label{tab:budget}
  \footnotesize
  \setlength{\tabcolsep}{2.4pt}
  \begin{tabular}{l|ccccc}
    \toprule
    & \textbf{AlphaEv.} & \textbf{DeepEv.} & \textbf{Self-Ev.} & \textbf{First-of-3} & \textbf{RecRSI}\\
    \midrule
    Rounds & 20 & 11 & 15 & 15 & 8\\
    NDCG@10 & 0.0531 & 0.0499 & \underline{0.0572} & 0.0567 & \textbf{0.0622}\\
    HR@10 & 0.0951 & 0.0877 & \underline{0.0991} & 0.0945 & \textbf{0.1024}\\
    \bottomrule
  \end{tabular}
\end{table}

\textbf{The advantage holds under an equal GPU budget.}
We also give every method one GPU hour on SASRec and CDs and let it run as many rounds as fit (Table~\ref{tab:budget}; protocol in Appendix~\ref{apdx:budget}).
RecRSI still improves Test NDCG@10 over AlphaEvolve, DeepEvolve and Self-EvolveRec by 17.1\%, 24.6\% and 8.7\% (all significant after Bonferroni correction), and AlphaEvolve, even with 20 rounds, stays below the five-round RecRSI of Table~\ref{tab:main} (0.0531 vs.\ 0.0594): additional untested retrainings do not close the gap. All five methods use 59.7--59.9 GPU minutes, including trials and evaluations, so this comparison accounts for the cost of obtaining early ranking feedback. RecRSI completes eight rounds, compared with 11--20 for the outcome-reward agents and 15 for First-of-3, yet achieves the highest NDCG@10 and HR@10. The result connects the quality gain to how retrainings are allocated: the alternatives can spend the time saved by omitting trials on further rounds, whereas RecRSI uses that time to compare edits by full-ranking evaluation.

\subsection{(RQ2) Comparing Selection Signals}
\label{sec:rq2}
\begin{figure}[t]
  \centering
  \includegraphics[width=\columnwidth]{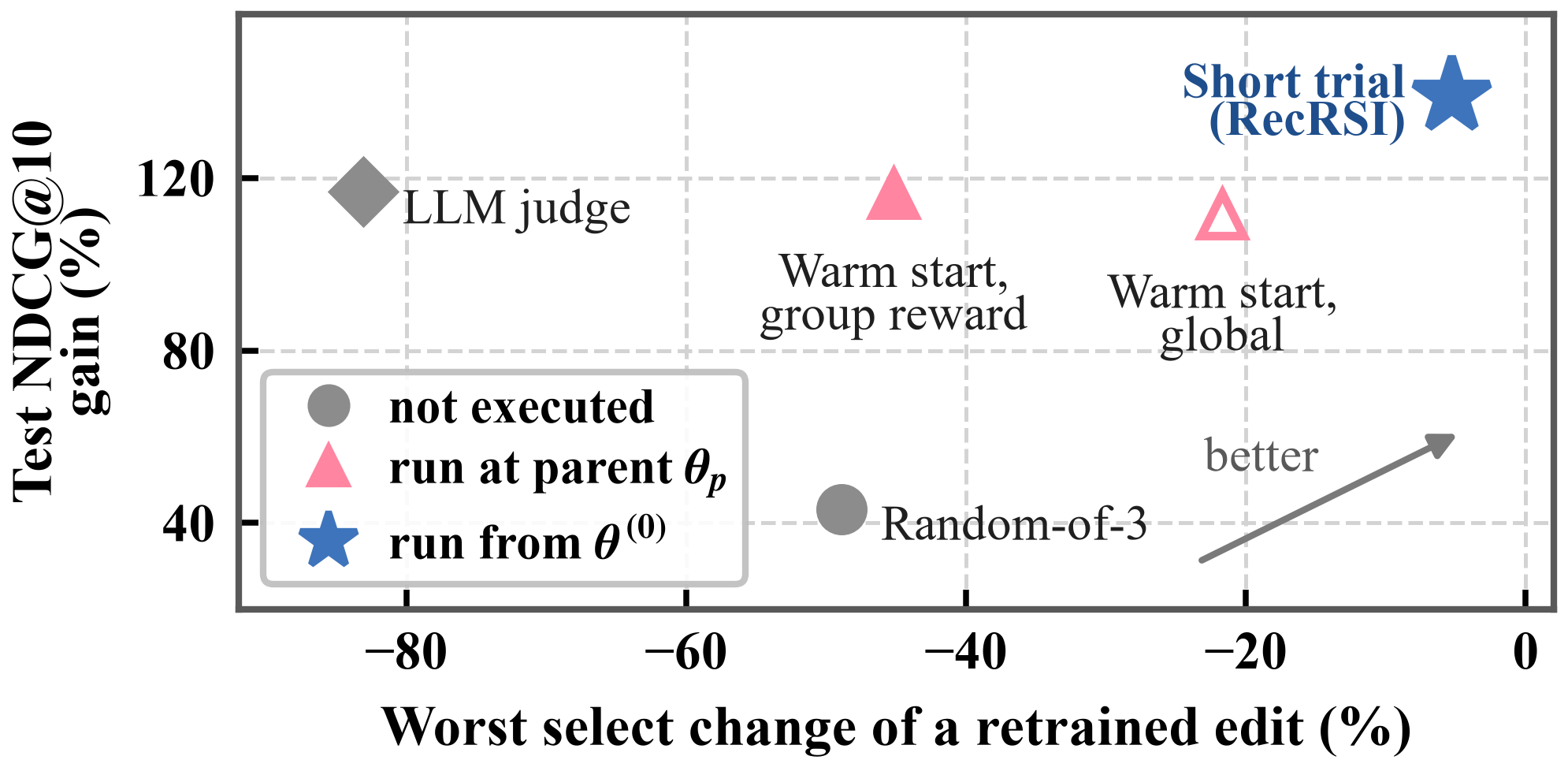}
  \vspace{-16pt}
  \caption{\textbf{Quality and risk of the signals that choose among three proposed edits} (SASRec): Test NDCG@10 gain over the initial model (averaged over CDs and Office) against the largest select drop of a retrained edit. Only the short trial is both the strongest and the safest.}
  \label{fig:risk}
  \Description{Scatter plot with five points. Random-of-3: worst change minus 48.9 percent, gain 43.2 percent. LLM judge: minus 83.1 percent, 116.9 percent. Warm start with group reward: minus 45.2 percent, 116.3 percent. Warm start global: minus 21.7 percent, 111.5 percent. Short trial (RecRSI): minus 5.3 percent, 139.2 percent, in the upper right corner. An arrow labelled better points to the upper right.}
  \vspace{-8pt}
\end{figure}

We next test whether the gain comes from the short-trial signal rather than from proposing three edits, and keep the proposer, the batch size and the retraining budget of RecRSI on SASRec and replace only the choice of Equation~\ref{eq:select}: \textbf{Random-of-3} picks a valid edit uniformly at random; the \textbf{LLM judge} ranks the edits from their diffs, hypotheses and the parent diagnosis without executing them, in the spirit of SIFT~\cite{fu2026sift}; \textbf{Warm start, global} continues each edit and the parent code for 1,500 steps from the trained parent $\theta_p$ rather than $\theta^{(0)}$, and ranks edits by the change of a smoothed full-ranking NDCG on probe users; and \textbf{Warm start, group reward} adds to this change a bonus on the rank group that each edit declared as its target and a penalty on the group it declared at risk, breaks ties by a gradient response at $\theta_p$, and may request one more batch when no edit scores positive.
Since these trajectories evolve different parents, their candidate sets differ after the first round (Table~\ref{tab:signal}, Figure~\ref{fig:risk}):

\begin{table}[t]
  \centering
  \caption{Choosing among three proposed edits with different signals (SASRec, Test). Worst: largest select NDCG@10 drop of a retrained edit. GPU: worker hours of both trajectories.}
  \label{tab:signal}
  \footnotesize
  \setlength{\tabcolsep}{2.6pt}
  \begin{tabular}{l|cc|cc|rr}
    \toprule
    & \multicolumn{2}{c|}{\textbf{CDs}} & \multicolumn{2}{c|}{\textbf{Office}} & & \\
    \textbf{Signal} & \textbf{N@10} & \textbf{H@10} & \textbf{N@10} & \textbf{H@10} & \textbf{Worst} & \textbf{GPU}\\
    \midrule
    Random-of-3 & 0.0470 & 0.0837 & 0.0119 & 0.0199 & $-$48.9\% & 1.61\\
    LLM judge & \textbf{0.0608} & \textbf{0.1005} & 0.0204 & 0.0342 & $-$83.1\% & 1.78\\
    Warm start, global & 0.0588 & \underline{0.1003} & 0.0200 & 0.0340 & $-$21.7\% & 2.33\\
    Warm start, group reward & 0.0552 & 0.0922 & \underline{0.0216} & \underline{0.0361} & $-$45.2\% & 2.70\\
    \midrule
    Short trial (RecRSI) & \underline{0.0594} & 0.0973 & \textbf{0.0242} & \textbf{0.0404} & $-$5.3\% & 2.44\\
    \bottomrule
  \end{tabular}
  \vspace{-8pt}
\end{table}

\textbf{Random choice among the same proposals loses most of the gain.}
Specifically, RecRSI improves Test NDCG@10 over Random-of-3 by 26.3\% on CDs and 103.4\% on Office (both significant), while Random-of-3 promotes no edit in the first round of either trajectory and once retrains an edit that nearly halves the select NDCG@10 of its parent.
Therefore, proposing several edits is not sufficient by itself: the edits differ widely in quality (Section~\ref{sec:rq3}), and the gain requires a signal that tells them apart.

\begin{figure}[b]
  \centering
  \includegraphics[width=\columnwidth]{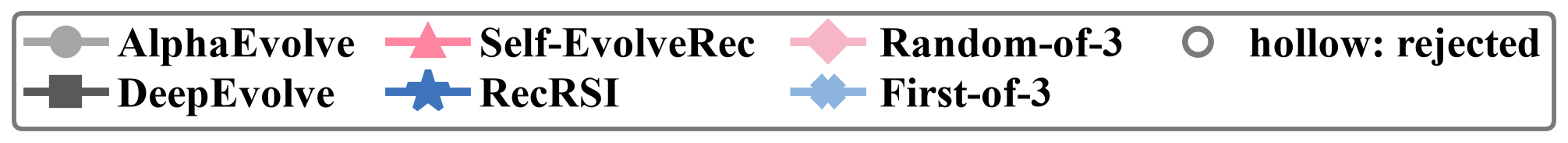}\\[-1pt]
  \begin{subfigure}[t]{0.49\columnwidth}
    \centering
    \includegraphics[width=\linewidth]{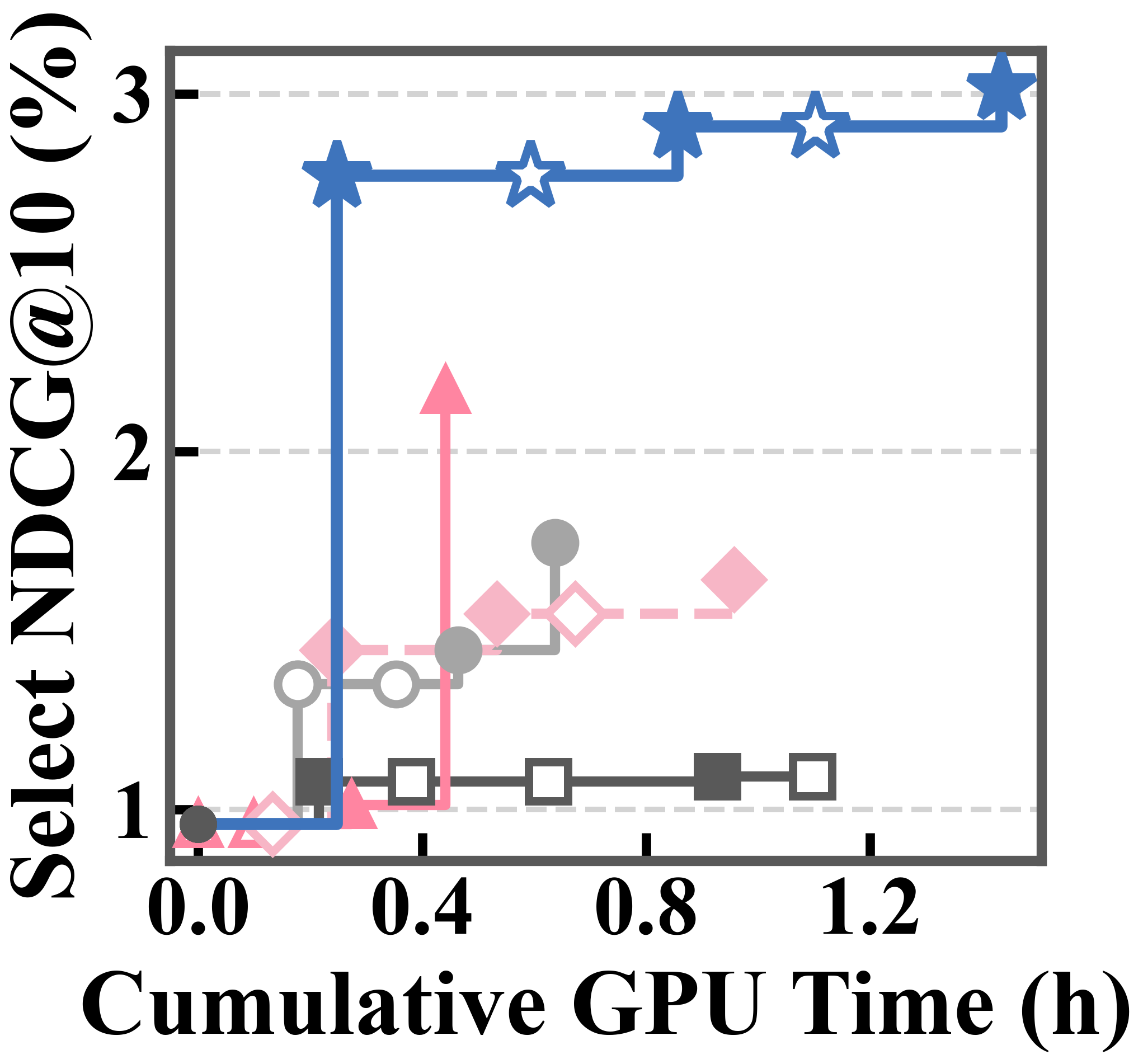}
    \caption{Office, five rounds}
    \label{fig:anytime-a}
  \end{subfigure}\hfill
  \begin{subfigure}[t]{0.49\columnwidth}
    \centering
    \includegraphics[width=\linewidth]{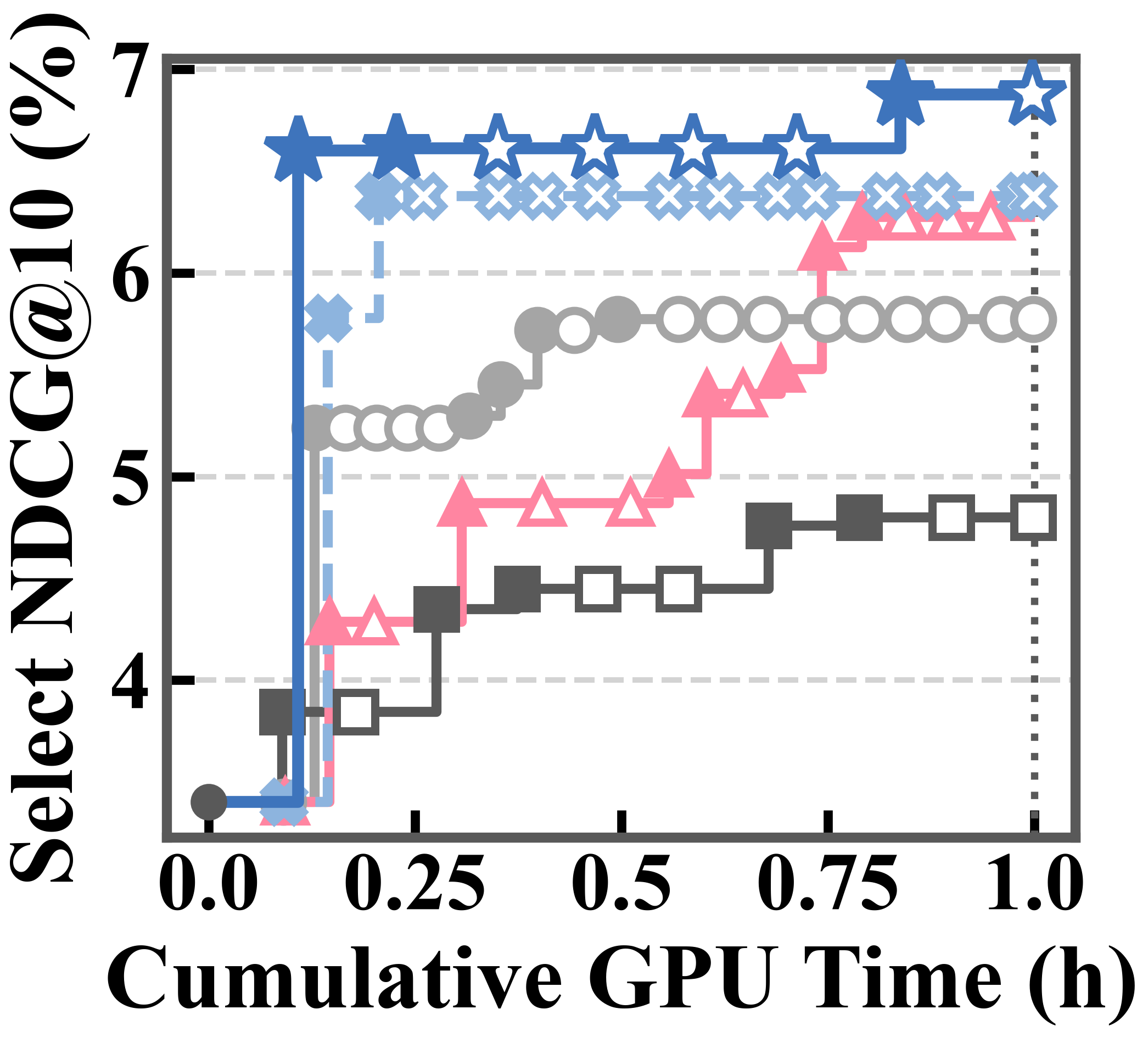}
    \caption{CDs, one GPU hour}
    \label{fig:anytime-b}
  \end{subfigure}
  \vspace{-8pt}
  \caption{\textbf{Select NDCG@10 vs.\ cumulative GPU time} (SASRec; hollow: rejected). After one round, RecRSI exceeds the (a) five-round and (b) one-hour results of all others.}
  \label{fig:anytime}
  \Description{Two step plots of select NDCG@10 versus cumulative GPU hours on SASRec. Panel a, Office with five rounds: AlphaEvolve, DeepEvolve, Self-EvolveRec and Random-of-3 stay between about 1 and 2.4 percent, while RecRSI jumps to about 2.7 percent after its first round at about 0.25 GPU hours and ends near 3 percent. Panel b, CDs with a budget of one GPU hour: RecRSI reaches 6.6 percent after its first round at 0.11 GPU hours and ends at 6.9 percent after 8 rounds; First-of-3 stays at 6.4 percent from 0.21 hours on over 15 rounds, Self-EvolveRec rises gradually to 6.3 percent over 15 rounds, AlphaEvolve to 5.8 percent over 20 rounds and DeepEvolve to 4.8 percent over 11 rounds.}
\end{figure}

\textbf{Skipping the trial loses both quality and safety.}
With a single edit per round ($K=1$), retrained without a trial as in the outcome-reward loop, the same proposer reaches a Test NDCG@10 of 0.0215 on SASRec Office, 11.1\% below RecRSI (significant), although it spends 1.94 instead of 1.44 GPU hours, and in its second round it retrains an edit that loses 93.3\% of the parent's select NDCG@10.
Likewise, under one GPU hour on CDs (Table~\ref{tab:budget}), \textbf{First-of-3}, which retrains the first valid edit of the same batch without a trial, runs more rounds but promotes no edit after its third (Figure~\ref{fig:anytime}(b)) and ends 9.7\% below RecRSI (significant).

\textbf{Short trials are the most reliable signal across datasets.}
On the one hand, the LLM judge is slightly higher than RecRSI on CDs (0.0608 vs.\ 0.0594 NDCG@10, not significant), but 15.7\% lower on Office, where it spends its first two retrainings on rejected edits, one of which loses 83.1\% of the parent's select NDCG@10.
On the other hand, both warm-start signals are below RecRSI on Office and in CDs NDCG@10, although they spend comparable or more GPU time (2.33 and 2.70 h vs.\ 2.44 h); on Office, all signals are significantly below RecRSI.
Overall, averaged over the two datasets, RecRSI improves the initial Test NDCG@10 by 139.2\%, compared with 116.9\% for the LLM judge, 111.5\% and 116.3\% for the two warm starts, and 43.2\% for random choice.
Moreover, RecRSI never retrains an edit that loses more than 5.3\% of its parent's select NDCG@10, whereas each of the other signals retrains at least one edit that loses over 20\%, so that only the short trial is both the strongest and the safest signal (Figure~\ref{fig:risk}).
We attribute this to training consistency: an edit that looks harmless on a converged parent, or plausible in a diff, can still train poorly from scratch, and only the trial detects this before full retraining.

\textbf{Only executed signals track retraining outcomes.}
Recomputing every signal for 38 edits that the outcome-reward trajectories had already retrained (Appendix~\ref{apdx:retro}), every executed signal correlates positively with the select gain of the retraining, and the short trial reaches $\rho=0.46$ and an AUC of 0.71, whereas the LLM judge, which has neither training consistency nor ranking consistency, is negatively correlated ($\rho=-0.14$, AUC 0.36): the effect of an edit lies in the data-dependent behaviour of training, which an LLM cannot read from a patch.

\subsection{(RQ3) Evolution Dynamics and Cost}
\label{sec:rq3}
\begin{figure}[t]
  \centering
  \begin{subfigure}[t]{0.5\columnwidth}
    \centering
    \includegraphics[width=\linewidth]{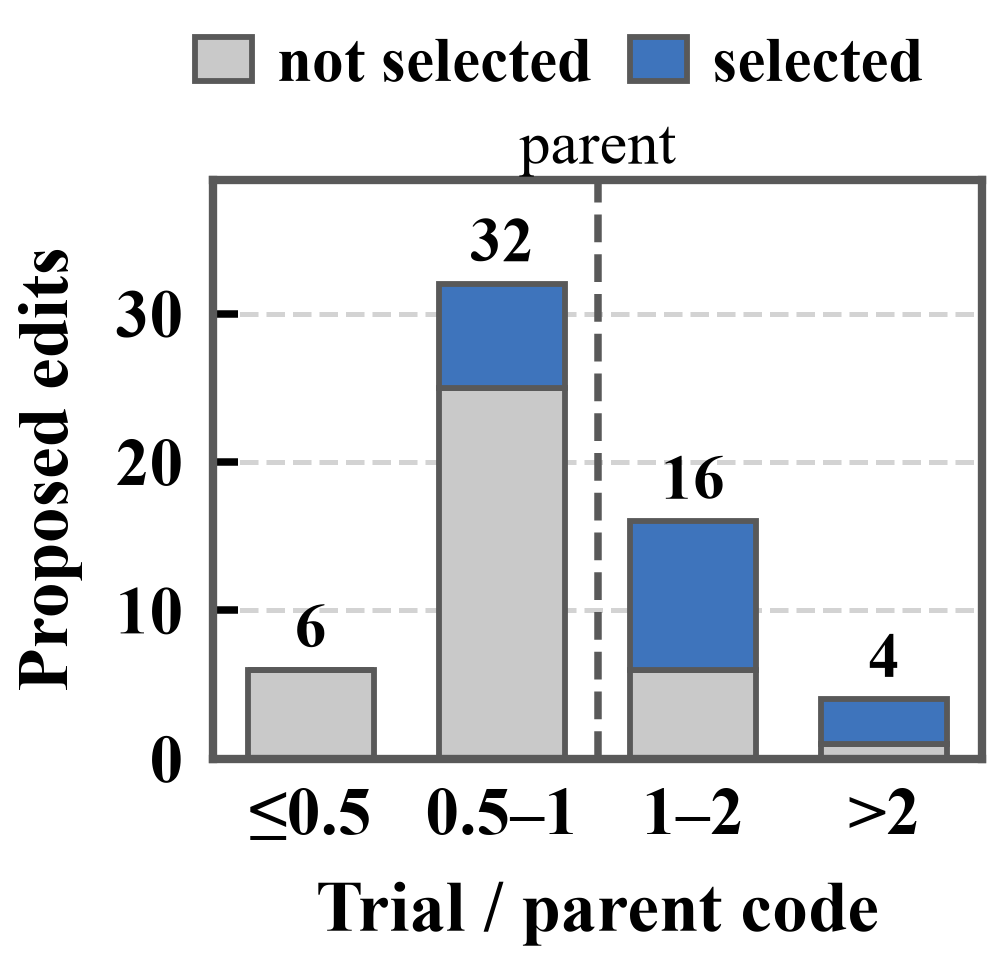}
    \caption{Trials of all proposed edits}
    \label{fig:spread-a}
  \end{subfigure}\hfill
  \begin{subfigure}[t]{0.49\columnwidth}
    \centering
    \includegraphics[width=\linewidth]{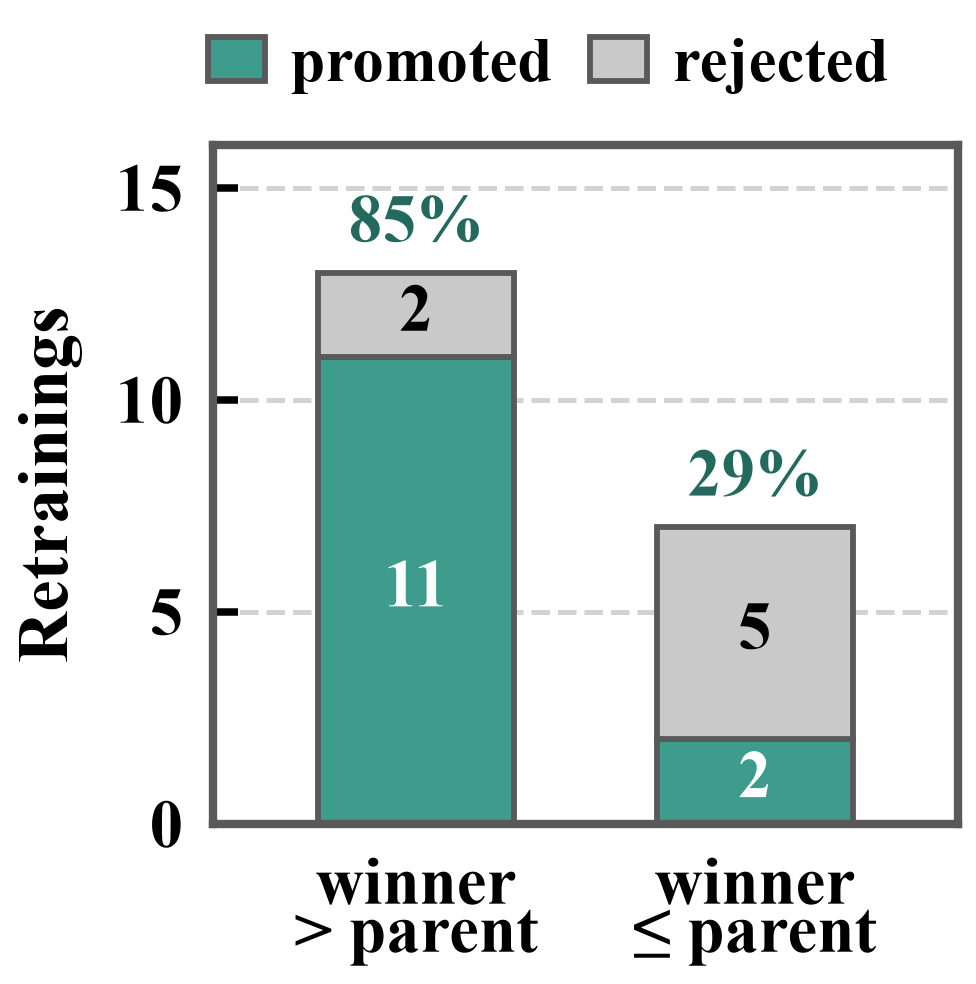}
    \caption{Outcome of the retraining}
    \label{fig:spread-b}
  \end{subfigure}
  \vspace{-8pt}
  \caption{\textbf{What the short trials reveal in the 20 rounds of RecRSI on CDs and Office.} (a) Trial NDCG@10 of the 58 proposed edits relative to the unchanged parent code (dashed line; blue: selected): most edits do not beat the parent. (b) Retrainings are mostly promoted when the selected edit beats the parent trial ($m_{a^{\star}}>m_p$).}
  \label{fig:spread}
  \Description{Panel a: stacked bars of proposed edits by trial ratio to the parent code: at most 0.5, 6 edits; 0.5 to 1, 32 edits (7 selected); 1 to 2, 16 edits (10 selected); above 2, 4 edits (3 selected); 38 of 58 lie left of the dashed parent line. Panel b: when the winner beats the parent, 11 of 13 retrainings are promoted (85 percent); otherwise 2 of 7 (29 percent).}
  \vspace{-8pt}
\end{figure}

Finally, Figures~\ref{fig:anytime} and~\ref{fig:spread} show how short trials change the course and the cost of evolution:

\textbf{Short trials front-load progress, and later rounds keep adding to it.}
Specifically, RecRSI promotes an edit in the first round of all six trajectories; after this round (0.09--0.29 GPU hours), its select NDCG@10 on Office already exceeds the five-round result of every outcome-reward baseline and of Random-of-3 (Figure~\ref{fig:anytime}(a)), and on CDs the one-hour result of every other method (Figure~\ref{fig:anytime}(b)).
Later rounds remain productive: 17 of its 24 retrainings in rounds 2--5 are promoted (Appendix~\ref{apdx:case-study} traces every round on LastFM). For example, SASRec on LastFM gains 31.7\% and 17.2\% select NDCG@10 in rounds 4 and 5 by training on random history crops, extending the earlier loss and sampler improvements with data augmentation.

\textbf{Proposals differ widely in quality, and the parent trial is an informative reference.}
Across the 20 rounds, 38 of the 58 proposed edits (65.5\%) do not exceed the parent trial score (Figure~\ref{fig:spread}(a)). The remaining 20 above-parent edits occur in 13 rounds; in the other seven rounds, even the highest-scoring candidate does not exceed the parent reference. The parent trial thus complements the comparison among candidates by indicating whether the best proposal improves early ranking over the unchanged recipe under the same training budget.
Moreover, when the selected edit beats the parent trial ($m_{a^{\star}}>m_p$), 11 of 13 retrainings are promoted; when it does not, only 2 of 7 are, with gains of 5.0\% and 2.9\% (Figure~\ref{fig:spread}(b)).
On LastFM, the selected edit beats the parent trial in 9 of the 10 rounds, all of which are promoted, and in HSTU round 3 the trials discard two edits that collapse (probe NDCG@10 of 0.0185 and 0), and the retraining goes to the third, which is promoted.
Skipping the retraining in rounds with $m_{a^{\star}}\le m_p$ would have saved five of the seven retrainings on CDs and Office that were not promoted, which makes $m_p$ an early-stopping rule for an evolution budget.

\section{Related Work}
\label{sec:related}
\textbf{Automated recommender development.}
Automated recommender design first searched predefined spaces, as in NASRec with its weight-sharing supernet~\cite{zhang2023nasrec}.
LLM agents remove this restriction by editing code: general systems such as FunSearch, AlphaEvolve and DeepEvolve evolve programs from execution scores~\cite{romera2024funsearch,novikov2025alphaevolve,liu2025deepevolve}, and recommender agents combine offline experiments with online validation~\cite{wang2026selfevolving}, accumulate experiment memory and skills~\cite{mu2026evorec}, route modification directions with a bandit~\cite{ling2026recharness}, or coordinate long-horizon research in industrial systems~\cite{yang2026agentx}.
Self-EvolveRec is closest to our setting: it diagnoses the parent numerically, simulates user critiques~\cite{zhang2024agent4rec}, and evolves its diagnostic code together with the model~\cite{kim2026selfevolverec}.
Beyond accuracy, \citet{xu2026retention} study what successive recommenders should retain during RSI.
These works choose edits to propose, whereas RecRSI selects edits for retraining and can use any of their proposers.

\textbf{Cheap evaluation for self-improvement and AutoML.}
Self-improving coding agents reduce the cost of judging a modification by evaluating it on task subsets~\cite{zhang2025dgm}, by allocating evaluations across lineages~\cite{wang2025hgm}, or by ranking patches with an LLM judge before any benchmark run~\cite{fu2026sift,zheng2023judge}; research preference models rank solutions or run pilot experiments first~\cite{foster2026rpm}, and AlphaEvolve passes programs through a cascade of increasingly expensive tests~\cite{novikov2025alphaevolve}.
In hyperparameter and architecture search, multi-fidelity methods train many configurations briefly and continue only the promising ones~\cite{jamieson2016successive,li2017hyperband,falkner2018bohb}, learning-curve extrapolation predicts final accuracy from a prefix~\cite{domhan2015speeding}, and zero-cost proxies score architectures at initialization~\cite{abdelfattah2021zerocost,mellor2021naswot}, a direction that Self-EvolveRec also offers as an efficient variant~\cite{kim2026selfevolverec}.
RecRSI borrows the low-fidelity trial of multi-fidelity search and can be read as a two-stage cascade, but its setting differs from both: multi-fidelity methods compare configurations that share one training objective, and cascade stages are tests designed by the user, whereas recommender code evolution compares open-ended LLM edits that often rewrite the objective itself, under a budget counted in full retrainings.
RecRSI designs this first stage for training code: it starts from the shared initialization and compares edits by the outcome's full-ranking evaluation.

\textbf{Sampled objectives in recommendation.}
Recommenders are commonly trained with sampled negatives or sampled softmax~\cite{yi2019sampling}, and both sampled metrics~\cite{krichene2020sampled} and sampled training objectives~\cite{petrov2023gsasrec,klenitskiy2023dross} are known to distort full-ranking evaluation.
Our measurements show that this distortion also hides from a self-improving agent which of its edits help, and RecRSI turns it into a design requirement for the selection signal.

\section{Conclusion}
\label{sec:conclusion}
Recursive self-improvement of recommender systems is held back because an edit's ranking quality is known only after a full retraining, which forces each retraining onto a single untested edit, while the signals available earlier, such as the sampled training loss, do not reflect full-ranking performance.
To this end, this work presented RecRSI, in which the LLM proposes a batch of diverse edits and short trials that follow the full training run and measure full-ranking performance choose the edit that receives the retraining.
Experiments with two backbones on three datasets showed that RecRSI significantly outperformed AlphaEvolve, DeepEvolve and Self-EvolveRec in all six settings at the same number of retrainings and avoided the regressions of the other selection signals.
The equal GPU-budget comparison further shows that selecting edits with ranking feedback improves the use of computation: RecRSI achieves higher Test NDCG@10 even when competing agents complete more rounds. Short trials therefore make ranking feedback actionable when choosing what to retrain.

\label{main-text-end}
\bibliographystyle{ACM-Reference-Format}
\bibliography{references}
\clearpage
\appendix

\section{Limitations}
\label{sec:limitations}
This work still has a few limitations.
On the one hand, the main comparison fixes five rounds per trajectory.
This horizon is close to where such searches already peak: Self-EvolveRec reaches its best model within 6--11 iterations in its own experiments~\cite{kim2026selfevolverec}, and under one GPU hour (Table~\ref{tab:budget}), the baselines run 11--20 rounds and RecRSI still leads; longer searches on larger datasets remain to be evaluated.
On the other hand, the large language model employed in this study is limited to a single open-weights model (GLM-5.3), and the effects of using different models have not yet been investigated.
Furthermore, whether the trial length established in the current experiments can be generalized to larger-scale datasets remains to be explored.

\section{Implementation Details}
\label{apdx:impl-details}
\textbf{Data.}
Table~\ref{tab:apdx-data} lists the statistics.
CDs is the processed version released with Self-EvolveRec and keeps its split; Office is the full official Amazon Reviews 2023 rating-only 5-core release without further filtering, ordered by timestamp with the source order breaking ties.
LastFM is built from the Last.fm-1K listening log at the level of songs; it keeps repeated listening events, removes only events that repeat the same user, timestamp and song, and is filtered iteratively on training interactions so that every song is heard by at least five users and every user hears at least five distinct songs.
Each LastFM user's last event is held out for Test and the second-to-last for validation, and Table~\ref{tab:apdx-data} counts only training-catalog items.

\textbf{Isolation of the evaluation sets.}
The split separates the evaluation sets at three levels.
\textit{Users}: the users with a validation interaction are assigned by a fixed hash of the user identifier to probe users (40\%), select users (40\%) and a reserved audit subset (20\%) that this paper does not use, so the probe and select sets share no user.
\textit{Interactions}: Test is not a third group of users but a later interaction of each user, held out for every user that has one (all users on Office and LastFM; the released Test users on CDs), so a probe or select user is also a Test user whenever it has a Test interaction, which is always the case on Office and LastFM.
A user's validation and Test interactions are distinct events, and the Test interaction is the later of the two.
\textit{Input histories}: training uses only the training interactions of all users; a probe or select user is scored on its validation interaction given its training sequence, and it is scored on its Test interaction given its training sequence followed by its validation interaction, which serves as context only and is never a training target.
The search reads only the training, probe and select files; the Test interactions are stored separately and are accessed once per finished trajectory, after the retained model has been sealed.
The item catalog and the item counts used by samplers are built from training interactions only.

\begin{table}[h]
  \centering
  \caption{Dataset statistics. Probe and Select: users whose validation interaction falls into each set (disjoint by user). Test: users with a held-out Test interaction, evaluated on an interaction later than their validation one; on Office and LastFM they comprise all users, including the probe and select users.}
  \label{tab:apdx-data}
  \small
  \setlength{\tabcolsep}{3.0pt}
  \begin{tabular}{l|rrr|rrr}
    \toprule
    \textbf{Dataset} & \textbf{Users} & \textbf{Items} & \textbf{Interactions} & \textbf{Probe} & \textbf{Select} & \textbf{Test}\\
    \midrule
    CDs & 14,335 & 11,436 & 126,225 & 4,043 & 4,013 & 9,716\\
    Office & 223,308 & 77,551 & 1,800,878 & 89,528 & 89,143 & 223,308\\
    LastFM & 985 & 175,635 & 12,785,061 & 388 & 398 & 985\\
    \bottomrule
  \end{tabular}
\end{table}

\textbf{Initial recommenders.}
Both backbones use hidden size 50, two layers, two attention heads, a maximum history of 50, dropout 0.2, Adam ($\beta=(0.9,0.98)$) with learning rate $10^{-3}$ and no weight decay, batch size 128, gradient clipping at 5, binary cross-entropy with one uniform negative per position, and exclusion of the user's history from negatives.
HSTU replaces the self-attention block of SASRec by the HSTU block (a joint projection to four SiLU-activated streams, pointwise SiLU attention normalized by the sequence length, and a layer-normalized output gated by one stream).
The agents may rewrite the model, the training step, and the sampler, but not the evaluator, the step cap, or the data access.
A short trial uses \(K_s = 4,000\) steps, about 15\% of the 26,000-step budget.

\textbf{Evaluation.}
Every evaluation follows the full-ranking protocol: it ranks all items of the training catalog and breaks ties by the lower item index.
Checkpoints are selected on the validation interactions of probe users during training; those of select users decide promotion; Test is evaluated once after each search.

\textbf{LLM interface.}
All methods use the same open-weight LLM (GLM-5.3) through the same client, with temperature 0.5 and at most 16,384 output tokens.
The prompt of RecRSI is a JSON document with the task, the interface contract, the rank and popularity groups, the parent's source code, the diagnosis, and the memory; the requested output is a list of edits, each with a name, a component, the motivating observation, a hypothesis, a mechanism, a target and a side-effect group, and search-and-replace patches.
The prompt omits fields whose names refer to select or Test values.

\section{Adaptation of the Baselines}
\label{apdx:adaptation}
The three outcome-reward frameworks are re-implemented in the shared recommender harness rather than run from their original code.
All of them use the same editable modules (model, training step, sampler, and diagnostic code), the same static and numerical execution checks, the same training setup, and the same promotion rule.
\textbf{AlphaEvolve} keeps a scored program archive with parent and inspiration sampling (25 programs, 10 elites, 5 islands; within five rounds, no migration between islands occurs).
\textbf{DeepEvolve} adds literature retrieval restricted to arXiv abstracts (at most two planned queries, one fallback and one follow-up query) and one reflection step.
\textbf{Self-EvolveRec} adds a numerical diagnosis and simulated user critiques before each edit and evolves its diagnostic code; when its prompt exceeds the context budget, earlier-round evidence and retrieved abstracts are compressed first, while the current source code and measurements are kept in full.

\section{Retrospective Signal Analysis}
\label{apdx:retro}
The 38 archived edits come from SASRec trajectories on CDs and Office of AlphaEvolve (7), DeepEvolve (9), Self-EvolveRec (6), and an earlier single-edit agent built on the same harness (16).
Each edit was fully retrained during its trajectory, and its outcome is the select gain over its parent; 20 edits have a gain above $\delta$.
Edits that share a parent form a group (18 groups, of which seven contain two to seven edits).
For each group, every signal is recomputed at the common parent with fixed seeds and probe users.
Since these edits were available while the selection signals were developed, the analysis is retrospective rather than a held-out test; the online comparison of Section~\ref{sec:rq2} provides the prospective evidence.
Table~\ref{tab:apdx-retro} reports the correlation with the outcome over all edits, the AUC of separating edits with a gain above $\delta$, and the number of multi-edit groups in which the signal ranks the best edit first; random choice would do so in 2.2 of the seven groups on average.
The raw gradient response at the parent is comparable to the short trial in correlation ($\rho=0.49$), but it is undefined for edits without differentiable scores, and its sign version, which reflects the direction of an Adam update~\cite{kingma2015adam}, drops to $\rho=0.23$.

\begin{table}[h]
  \centering
  \caption{Retrospective agreement of selection signals with retraining outcomes on 38 archived edits. Top-1: multi-edit parent groups (out of seven) in which the signal ranks the best retrained edit first (random choice: 2.2 expected).}
  \label{tab:apdx-retro}
  \small
  \setlength{\tabcolsep}{4pt}
  \begin{tabular}{l|rrrr}
    \toprule
    \textbf{Signal} & $n$ & $\rho$ & \textbf{AUC} & \textbf{Top-1}\\
    \midrule
    LLM judge (no execution) & 27 & $-$0.144 & 0.363 & 3/7\\
    Gradient response, sign & 37 & 0.229 & 0.603 & 4/7\\
    Gradient response, raw & 37 & 0.487 & 0.725 & 3/7\\
    Gradient, raw, ranks 11--100 & 37 & 0.297 & 0.703 & 2/7\\
    Warm start, group reward & 38 & 0.291 & 0.668 & 3/7\\
    Warm start, global (smoothed) & 38 & 0.421 & 0.707 & 3/7\\
    Warm start, global (NDCG@10) & 38 & 0.419 & 0.729 & 3/7\\
    Short trial (RecRSI) & 38 & 0.462 & 0.708 & 3/7\\
    \bottomrule
  \end{tabular}
\end{table}

\section{Case Study: Five Rounds on LastFM}
\label{apdx:case-study}
\begin{figure*}[t]
  \centering
  \includegraphics[width=\textwidth]{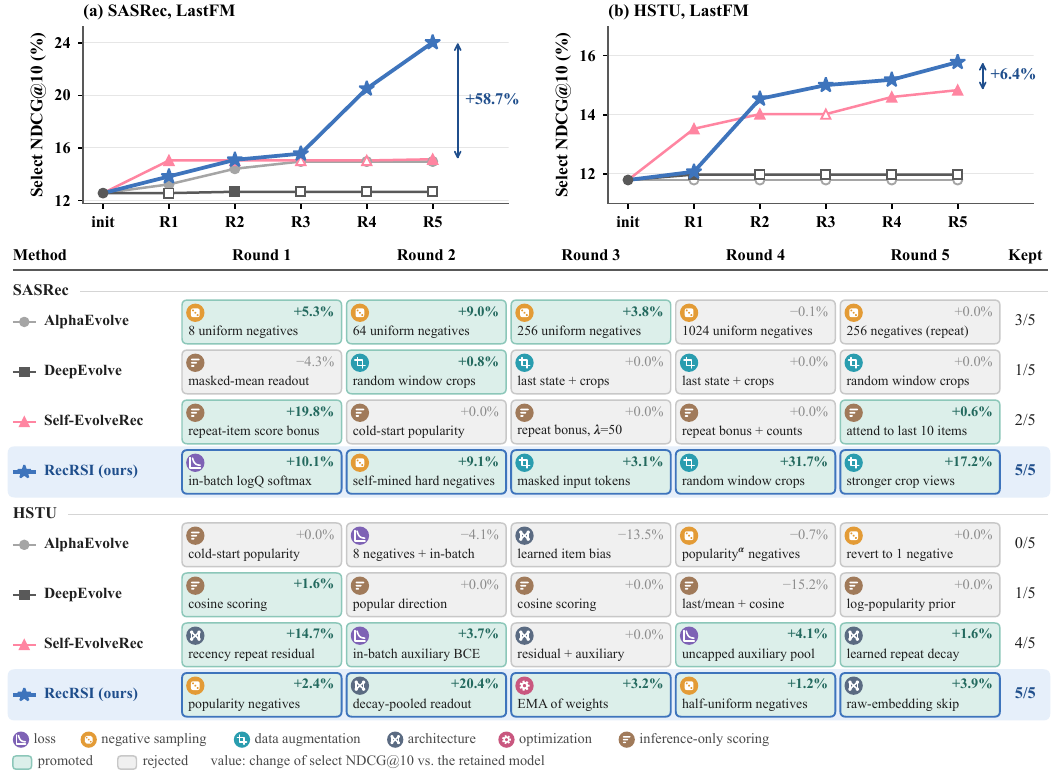}
  \vspace{-14pt}
  \caption{\textbf{Every round of every method on LastFM} (one full retraining per round). Top: select NDCG@10 of the retained model after each round (filled marker: promoted; arrow: gap to the strongest baseline after five rounds). Bottom: the retrained edit of each round with an icon for its type and the change of select NDCG@10 against the retained model (green: promoted; gray: rejected). Edit labels summarize the proposal of each round.}
  \label{fig:apdx-case}
  \Description{Two line charts of select NDCG@10 versus round for AlphaEvolve, DeepEvolve, Self-EvolveRec and RecRSI on LastFM: RecRSI ends 58.7 percent (SASRec) and 6.4 percent (HSTU) above the strongest baseline. Below, a grid with one cell per method, backbone and round naming the retrained edit, its type (loss, negative sampling, data augmentation, architecture, optimization, inference-only scoring), the select change, shaded green if promoted and gray if rejected. SASRec AlphaEvolve raises the number of uniform negatives in every round; Self-EvolveRec on SASRec and DeepEvolve on HSTU change only inference-time scoring; RecRSI retrains a softmax loss, hard negatives and three augmentation edits on SASRec with gains of 10.1, 9.1, 3.1, 31.7 and 17.2 percent, and five edits of three types on HSTU, all promoted.}
  \vspace{-6pt}
\end{figure*}
Figure~\ref{fig:apdx-case} follows all four methods through the five rounds on LastFM, the dataset for which the complete per-round record of every trajectory is part of our released results.

\textbf{Gains accumulate over the rounds.}
RecRSI is promoted in all ten rounds, and most of its progress comes after the first round: on SASRec, the largest gains come in rounds 4 and 5 (+31.7\% and +17.2\% select NDCG@10), from edits that train on random crops of the input history, and on HSTU in round 2 (+20.4\%), from a decay-pooled readout.
Self-EvolveRec is ahead after round 1 on both backbones, but RecRSI leads from round 2 on and ends 58.7\% (SASRec) and 6.4\% (HSTU) above the strongest baseline.

\textbf{The baselines spend retrainings on narrow or inert edits.}
With a single edit per round, each baseline tends to follow one line: AlphaEvolve on SASRec only raises the number of uniform negatives (from 8 to 1,024), and Self-EvolveRec on SASRec and DeepEvolve on HSTU change only how the trained model scores items, which leaves training untouched; such inference-only edits make up 12 of the 30 baseline retrainings, and three of them are promoted.
Several rounds also re-propose an earlier edit, such as the window crops of DeepEvolve in SASRec rounds 3--5, which reproduce the select NDCG@10 of round 2.
The edits that RecRSI retrains span five types, none of them inference-only.

\textbf{The trials filter the batch.}
In seven of the ten rounds, at least one proposed edit trains below the unchanged parent code in its trial, and in each of these rounds that has an edit above the parent, this edit receives the retraining instead.
In round 3 of HSTU, two of the three edits collapse in their trials (probe NDCG@10 of 0.0185 and 0) and the retraining goes to the third, the only selected edit on LastFM below the parent trial, which is promoted with a small gain (+3.2\%), as $m_p$ predicts (Section~\ref{sec:rq3}).

\section{Recorded Cost}
\label{apdx:cost}
All training and evaluation run on a single NVIDIA T4 GPU.
GPU times are worker wall-clock seconds accumulated over a trajectory, excluding the shared initial training and the final Test evaluation, and include probe evaluations during training, contract checks, diagnoses and group audits.
On Office, a single SASRec retraining takes about 11 minutes (median over our runs) and up to 30 minutes.
On CDs and Office, a single short trial takes a median of 81--103\,s, compared with a median of 236--788\,s for a full retraining, depending on the dataset and backbone.
The trials therefore do not buy the quality of RecRSI with more compute than its alternatives: on SASRec Office, a five-round RecRSI trajectory takes 1.44 GPU hours, less than the single-edit variant without trials (1.94\,h) and the two warm-start signals (1.74 and 1.89\,h), each of which ends with a lower Test NDCG@10 (Section~\ref{sec:rq2}).
When every method instead receives the same GPU budget, RecRSI remains ahead of all of them (Appendix~\ref{apdx:budget}).

\section{Equal GPU-Budget Comparison}
\label{apdx:budget}
\textbf{Protocol.}
On SASRec and CDs, AlphaEvolve, DeepEvolve, Self-EvolveRec, First-of-3 and RecRSI start from the same initial model, configuration, data split and seed as in Table~\ref{tab:main}, and each searches with its own mechanism until its remaining budget of 3,600 GPU seconds cannot fit another job.
The budget counts every GPU job of the search: diagnoses, short trials, full retrainings, failed jobs, and the evaluations on probe and select users; a training run that a method reuses from its cache is charged at the GPU time of its original run.
The shared initial training and the single final Test evaluation are outside the budget, and the numbers of LLM tokens are not equalized.
The fixed number of rounds and the candidate limit of the main comparison are lifted, so that a method with cheaper rounds runs more of them; the training setup, the promotion rule and all other settings are those of Section~\ref{sec:setup}.
First-of-3 uses the proposer, the batch size of $K=3$, the contract checks and the repair request of RecRSI, and retrains the first valid edit of the batch without a trial.
All five use 59.7--59.9 GPU minutes.

\textbf{Significance.}
Table~\ref{tab:apdx-budget} reports the paired bootstrap over the 9,716 Test users (5,000 resamples).
The intervals are taken at level 99.8\%, which is stricter than the Bonferroni level $1-0.05/4$ (98.75\%) for the four comparisons, and all four lie above zero.

\begin{table}[h]
  \centering
  \caption{Paired bootstrap of the Test NDCG@10 difference between RecRSI and each method under the equal GPU budget (SASRec, CDs; 99.8\% intervals).}
  \label{tab:apdx-budget}
  \small
  \setlength{\tabcolsep}{4pt}
  \begin{tabular}{l|rrr}
    \toprule
    \textbf{RecRSI vs.} & \textbf{$\Delta$NDCG@10} & \textbf{Interval} & \textbf{Rel.}\\
    \midrule
    AlphaEvolve & 0.0091 & [0.0038, 0.0149] & +17.1\%\\
    DeepEvolve & 0.0123 & [0.0068, 0.0178] & +24.6\%\\
    Self-EvolveRec & 0.0050 & [0.0003, 0.0101] & +8.7\%\\
    First-of-3 & 0.0055 & [0.0008, 0.0105] & +9.7\%\\
    \bottomrule
  \end{tabular}
\end{table}

\end{document}